\documentclass[a4paper,12pt]{article}
\pdfoutput=1

\usepackage[a4paper,left=80pt,right=80pt,top=70pt,bottom=71pt]{geometry}
\usepackage[latin1]{inputenc}
\usepackage{dsfont}
\usepackage{amsfonts,amsmath,amssymb}
\usepackage{amsthm}
\usepackage{graphicx}
\usepackage[small,bf]{caption}
\usepackage{subcaption}
\usepackage{booktabs}
\usepackage[numbers,sort&compress]{natbib}
\allowdisplaybreaks[1]
\usepackage{color}
\usepackage{ulem}
\usepackage{tikz}
\usepackage{multirow}
\usetikzlibrary{arrows}
\usepackage{xspace}
\usepackage{cancel}
\usepackage[latin1]{inputenc}
\usepackage[T1]{fontenc}
\usepackage{graphicx,amssymb,amsfonts,amsmath,slashed}
\usepackage{feynmp-auto}
\usepackage{cancel}
\usepackage{enumitem}
\def\bal#1\eal{\begin{align}#1\end{align}}
\newcommand{\be}{\begin{equation}}
\newcommand{\ee}{\end{equation}}
\newcommand{\bea}{\begin{eqnarray}}
\newcommand{\eea}{\end{eqnarray}}
\newcommand{\besub}{\begin{subequations}}
\newcommand{\eesub}{\end{subequations}}
\newcommand{\ba}{\begin{array}}
\newcommand{\ea}{\end{array}}
\newcommand{\bi}{\begin{itemize}}
\newcommand{\ei}{\end{itemize}}
\newcommand{\nn}{\nonumber}

\newcommand{\BR}{{\rm BR}}

\newcommand{\Lcal}{{\cal L}}
\newcommand{\hc}{{\textrm{h.c.}}}

\begin{document}

\begin{titlepage}

\flushright{ 
 }
 
\vspace*{1.5cm}

\begin{center}
{\Large 
{\bf
Multi--Higgs Boson Probes of the Dark Sector
}}
\\
[1.2cm]

{
{\bf
Marvin Flores$^{1,2}$, Christian Gross$^{3}$, Jong Soo Kim$^{4}$,\\ Oleg Lebedev$^{5}$ and Subhadeep Mondal$^{5}$
}}
\end{center}

\vspace*{1.2cm}

\centering{
$^{1}$ 
\it{School of Physics, University of the Witwatersrand, \\ Johannesburg, South Africa
}

\vspace*{0.1cm}
$^{2}$ 
\it{National Institute of Physics, University of the Philippines, \\ Diliman, Quezon City, Philippines
}

\vspace*{0.1cm}
$^{3}$ 
\it{Dipartimento di Fisica dell'Universit{\`a} di Pisa and INFN, Sezione di Pisa,\\ IT-56127 Pisa, Italy
}

\vspace*{0.1cm}
$^{4}$ 
\it{National Institute for Theoretical Physics and School of Physics, \\ University of the Witwatersrand, Johannesburg, South Africa
}

\vspace*{0.1cm}
$^{5}$ 
\it{Department of Physics and Helsinki Institute of Physics, \\
Gustaf H\"allstr\"omin katu 2a, FI-00014 Helsinki, Finland
}}	

\vspace*{1cm}

\begin{abstract}
\noindent
 {We consider dark sectors with spontaneously broken gauge symmetries, where} cascade decays of the dark sector fields naturally produce multi--Higgs boson final states along with dark matter. Our  study focuses on two and three Higgs boson final states with missing energy using a multivariate analysis with Boosted Decision Trees. We find that the di--Higgs boson channel is quite promising for the $\bar b b + \gamma \gamma$ and $\bar b b + \bar \ell \ell$ decay modes. The tri--Higgs boson final state with missing energy, on the other hand, appears to be beyond the reach of the LHC in analogous channels. This may change when fully hadronic Higgs boson decays are considered.
\end{abstract}


\end{titlepage}
\newpage

\tableofcontents

\section{Introduction}

The scalar sector of the Standard Model (SM) remains relatively little explored compared to the gauge sector. In particular, 
 no Higgs boson self-interaction has been measured. 
 It is a well motivated and experimentally viable possibility that the Higgs field plays the role of a portal into the hidden sector which may contain dark 
 matter and possibly other cosmologically relevant fields~\cite{Silveira:1985rk,Schabinger:2005ei,Patt:2006fw}. Exploring this portal requires Higgs coupling measurements as well as a search for extra states.
 The Higgs self--coupling can be determined 
 via Higgs boson pair production~-- one of the central objectives of the future (HL)-LHC searches, cf.~eg.~\cite{DiMicco:2019ngk} for a recent summary.
 Higgs portal dark matter can manifest itself in di--Higgs or multi--Higgs final states with large missing energy, which appears due to emission of undetected dark matter.
 Analogous signals arise in a multitude of New Physics models as studied in
 \cite{Matchev:1999ft,Kang:2015nga,Kang:2015uoc,Brivio:2017ije,Arganda:2017wjh,Chen:2018dyq,Bernreuther:2018nat,Titterton:2018pba,Biswas:2016ffy,Han:2015pwa,Papaefstathiou:2015paa,Etesami:2015caa,Banerjee:2016nzb,Banerjee:2019jys,Adhikary:2018ise,Banerjee:2018yxy}.
LHC searches for di-Higgs + \cancel{$E$}$_T$ have been presented in~\cite{Sirunyan:2017bsh,Sirunyan:2017obz,Aaboud:2018htj}.

Within a similar framework, a study of di-Higgs + \cancel{$E$}$_T$ production with the 4$b$ + \cancel{$E$}$_T$ final state has been performed in~\cite{Blanke:2019hpe}.
It utilises a jet substructure technique to reconstruct the boosted Higgs bosons. While for heavy dark states this method is efficient, in the intermediate mass range 
(below 700 GeV or so) it is less reliable.
 In our work, we consider both 2 and 3 Higgs final states, which subsequently decay into
$\bar b b$, $\gamma\gamma$ and $WW$. These have the advantage of being cleaner channels with a lower background. Instead of using a traditional cut--based analysis, we employ (as done for 4$b$ + \cancel{$E$}$_T$ in~\cite{Blanke:2019hpe}) a multivariate technique with Boosted Decision Trees which leads to much better sensitivity to New Physics. 
Further probes of the model are provided by monojet + \cancel{$E$}$_T$ searches \cite{Kim:2015hda}. 
The analysis of \cite{Kim:2015hda} is based on rectangular cuts and as such leads to lower sensitivity than the present study does.

 To set the framework for our study, we introduce a simplified model in which cascade decays of heavier states produce dark matter and the Higgses. This model is motivated by non--Abelian gauged hidden sectors in which gauge symmetry is broken completely by VEVs of a minimal set of dark Higgs multiplets \cite{Gross:2015cwa}. Such models automatically lead to stable dark matter
 which consists of a subset of gauge fields (and possibly extra scalars). Smaller groups like U(1) and SU(2) \cite{Hambye:2008bq, Lebedev:2011iq} do not allow for the needed cascade decays, while
 SU(3) and larger groups have all the necessary features. Further options, e.g. when part of the gauge group condenses, have been explored in~\cite{Buttazzo:2019mvl,Buttazzo:2019iwr}.

This paper is structured as follows.
In Section 2, we motivate our study and introduce the simplified model.
In Section 3, we perform an LHC study of the di--Higgs final state with missing energy using multivariate analysis with Boosted Decision Trees and present the resulting sensitivity to model parameters.
In Section 4, our findings on the tri--Higgs final state with missing energy are summarized.
Section 5 concludes our study.

\section{Motivation: multi--Higgs states and missing energy from dark gauge sectors}

\subsection{Dark Higgsed gauge sectors}

The presence of stable states which can play the role of dark matter is a common feature of dark sectors with spontaneously broken gauge symmetry.
In particular, breaking SU(N) completely with the minimal number of dark Higgs fields automatically leads to stable dark matter.
To be specific, let us briefly summarize the main relevant features of the SU(3)
example~\cite{Gross:2015cwa} (see also \cite{Karam:2016rsz,Poulin:2018kap}). The set--up contains two dark Higgs triplets $\phi_i$ to break the symmetry completely.\footnote{{This is the minimal scalar field content needed to break SU(3) completely.
With fewer scalars, part of the gauge group would remain unbroken and may eventually condense, leading to  different phenomenology, cf.~\cite{Buttazzo:2019mvl,Buttazzo:2019iwr}.}}
The Lagrangian is
$
\Lcal_{\rm SM} + \Lcal_{\rm portal} + \Lcal_{\rm hidden} \;,
$
where 
\besub
\bal
-\Lcal_{\rm SM} &\supset V_{\rm SM} =\frac{\lambda_{}}{2} |H|^4+m_{}^2 |H|^2 \;,
 \\
-\Lcal_{\rm portal} &= 
|H|^2\left( \sum_{i=1}^2 \lambda_{Hii} |\phi_i|^2- \left( \lambda_{H12}\phi_1^\dagger \phi_2 + \hc \right) \right)\;,
 \\
\Lcal_{\rm hidden} &= - \frac12 \textrm{tr} \{G_{\mu \nu} G^{\mu \nu}\} + \sum_{i=1}^2 |D_\mu \phi_i|^2 -V_{\rm hidden} \,.
\eal
\eesub
Here, $G_{\mu \nu}=\partial_\mu A_\nu - \partial_\nu A_\mu + i \tilde g [A_\mu,A_\nu]$ is the field strength tensor of the SU(3) gauge fields $A_\mu^a$ with gauge coupling $\tilde g$, $D_\mu \phi_i = \partial_\mu \phi_i + i \tilde g A_{\mu} \phi_i$, and $H$ is the Higgs doublet, which in the unitary gauge can be written as $H^T= (0,v+h)/\sqrt{2}$.
The potential for the dark Higgses is given by 
\bal
V_{\rm hidden}(\phi_1,\phi_2) &=
m_{11}^2 |\phi_1|^2
+ m_{22}^2 |\phi_2|^2
- ( m_{12}^2 \phi_1^\dagger \phi_2 + \hc )
\nn \\ 
& 
+ \frac{\lambda_1}{2} |\phi_1|^4
+ \frac{\lambda_2}{2} |\phi_2|^4
+ \lambda_3 |\phi_1|^2 |\phi_2|^2
+ \lambda_4 | \phi_1^\dagger\phi_2 |^2
\nn \\ 
& 
+ \left[
\frac{ \lambda_5}{2} ( \phi_1^\dagger\phi_2 )^2
+ \lambda_6 |\phi_1|^2
( \phi_1^\dagger\phi_2)
+ \lambda_7 |\phi_2|^2
( \phi_1^\dagger\phi_2 )
+ \hc \right] \,.
\label{V}
\eal
In the unitary gauge, $\phi_1,\phi_2$ can be written as
\be \label{unitarygauge}
\phi_1={1\over \sqrt{2}} \,
\left( \begin{array}{c}
0\\0\\v_1+\varphi_1
\end{array} \right) \,,
\quad 
\phi_2= {1 \over \sqrt{2}}\,
\left( \begin{array}{c}
0\\v_2+\varphi_2\\v_3+\varphi_3 + i (v_4 +\varphi_4)
\end{array} \right) ~,
\ee
where the $v_i$ are real VEVs and $\varphi_{i}$ are real scalar fields. 
 \begin{table}[t]
 \begin{center}
 \begin{tabular}{|c|c|}
\hline
 fields & $Z_2 \times Z_2'$
 \\ \hline \hline 
$h, \varphi^1, \varphi^2, \varphi^3, A_\mu^7$ & $(+,+)$
\\ \hline 
$A_\mu^2,A_\mu^5$& $(-,+)$
\\ \hline 
$A_\mu^1,A_\mu^4$& $(-,-)$
\\ \hline 
$\varphi^4,A_\mu^3,A_\mu^6, A_\mu^8$& $(+,-)$
\\ \hline
\end{tabular}
\end{center}
\vspace*{-10pt}
\caption{\label{parities} $Z_2 \times Z_2'$ parities of the scalars and dark gauge bosons.
}
 \end{table}

For simplicity, we assume an unbroken $CP$ symmetry in the scalar sector, i.e. we assume that the couplings are real and $v_4=0$.\footnote{If this assumption is relaxed, the model still has dark matter which consists of a subset of stable states considered here \cite{Arcadi:2016qoz}.}
As explained in~\cite{Gross:2015cwa,Arcadi:2016kmk}, the model then has an unbroken U(1)$\times Z'_2$ global symmetry.
Its $Z_2 \times Z'_2$ subgroup leads to stability of multicomponent DM.
The parities of the fields under $Z_2 \times Z'_2$ transformations are summarized in Table~\ref{parities}.

The lightest states with non--trivial parities cannot decay to the Standard Model particles and are viable DM candidates.
 The hidden vector and scalar fields can mix with other fields of the same $Z_2 \times Z_2'$ parity.

 In this example, $A_\mu^{1-3}$ and $\varphi^4$ play the role of dark matter. The other vectors $A_\mu^{4-8}$ are heavier and decay into $A_\mu^{1-3}$ with scalar emission.
 The scalar couplings to the vectors depend on the VEVs of the triplets as well as their mixing with the SM Higgs.
 The CP-even scalars are all expected to mix and their mass eigenstates include 
 the 125 GeV Higgs $h$, a heavier scalar $H$ and two more heavy scalars which play no significant role in our discussion. 
 Given that $H$--emission is often kinematically forbidden,
 cascade decays of the heavy states naturally produce multi--Higgs $h$
 signals.

\subsection{Simplified model}

In our study, only main features of this set--up play a role. Hence, it is convenient to introduce a simplified model which inherits salient features of the 
Higgs portal framework.
Consider an extension of the SM by three fields: a Higgs--like scalar $H$ with mass $m_{H}$, a stable vector field $A_L$ with mass $m_{A_L}$ 
and a heavier unstable vector field $A_H$ with mass $m_{A_H}$. Dark matter is assumed to be composed of $A_L$, while the 125 GeV Higgs $h$ 
and the heavier Higgs $H$ are mixtures of the SM Higgs doublet and the hidden sector singlet characterized by the mixing angle $\theta$.

The interactions of the prototype fields are listed in \cite{Gross:2015cwa}. 
The result depends on which of the triplets the SM Higgs mixes with predominantly. For our purposes, it is convenient to parametrize the couplings in terms of the mixing angles as follows.
The couplings of $h,H$ to SM matter and SM gauge fields are given by
\be
\Lcal \supset \left(h \cos \theta + H \sin \theta \right) \frac1v \left [ 2 m_W^2 W_\mu^+ W^{\mu -} + m_Z^2 Z_\mu Z^\mu - \sum_f m_f \bar f f \right] ,
\ee
while the couplings of $h,H$ to $A_L$, $A_H$ are given by
\bal 
\mathcal{L} & \supset 
\left(- h \sin \theta + H \cos \theta \right) \frac{ \tilde g }2
\left[ \cos \delta ~ m_{A_L} (A_L^\mu)^2 + \sin \delta ~ m_{A_H} (A_H^\mu)^2 \right] \,.
\eal
In addition to the masses, the New Physics parameters are: the dark gauge coupling $\tilde g$, the Higgs mixing angle $\theta$ and an angle $\delta$ that sets the relative strength of the scalar coupling to the dark vectors.\footnote{{Within the UV complete model of the previous subsection, $\sin \delta$ depends on the composition of the heavy Higgs $H$ in terms of the CP even dark scalars.}}
Further relevant couplings include the terms 
\begin{equation}
\mathcal{L} \supset -\kappa_{112} \; \frac v2 H h^2 - \sigma \, (A_L)_\mu \, A^\mu_H \, h\;.
\end{equation}
Here $\kappa_{112}$ is not fixed by the other model parameters, so that we can take $\BR(H\to hh)$ to be a free variable. The $\sigma$--term, which is {also  a free parameter within the simplified model}, accounts for decay
of the heavy gauge boson into dark matter and SM states.

\begin{figure}[t]
\begin{center}
\includegraphics[scale=0.365]{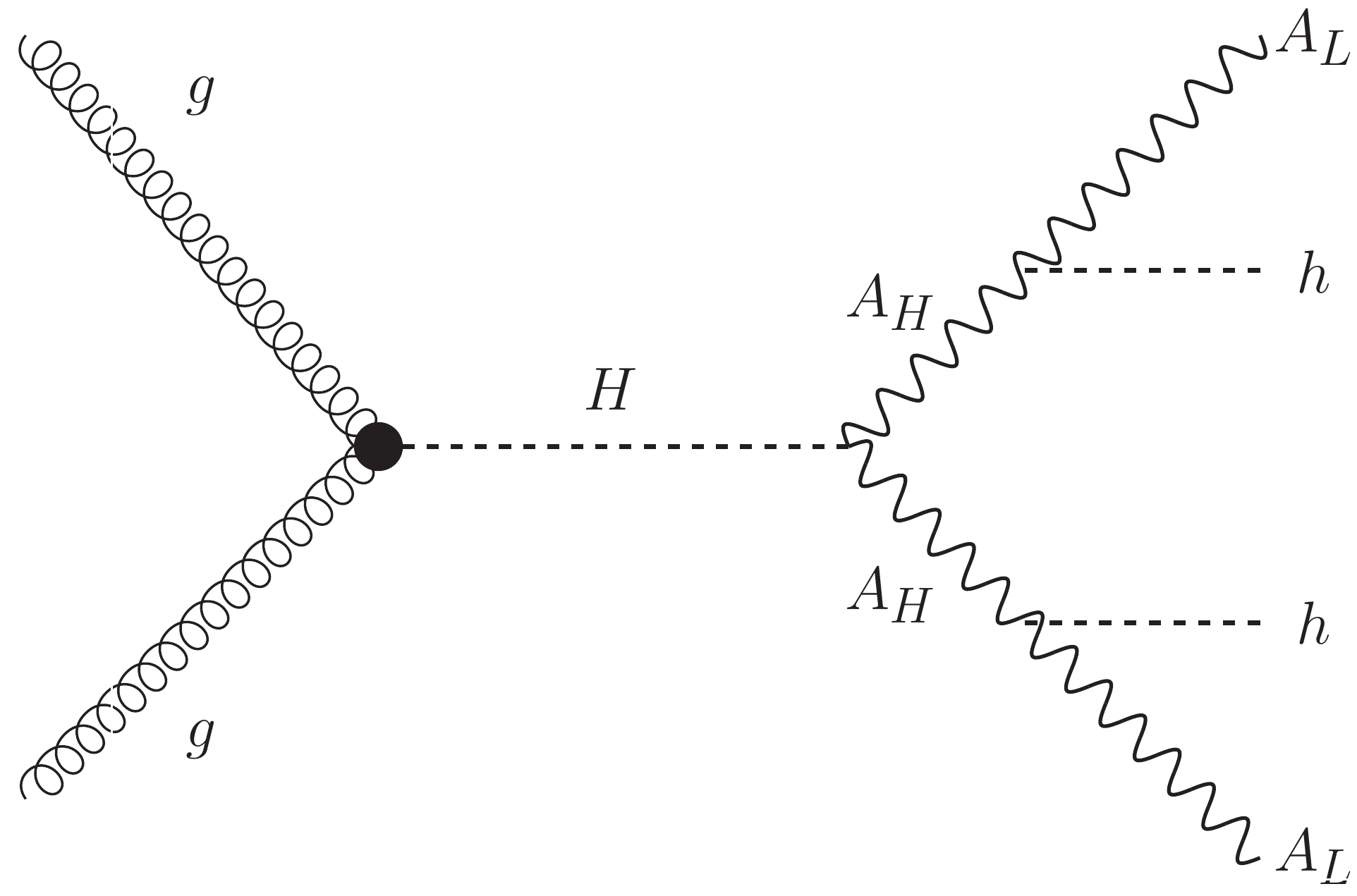} 
\qquad
\includegraphics[scale=0.365]{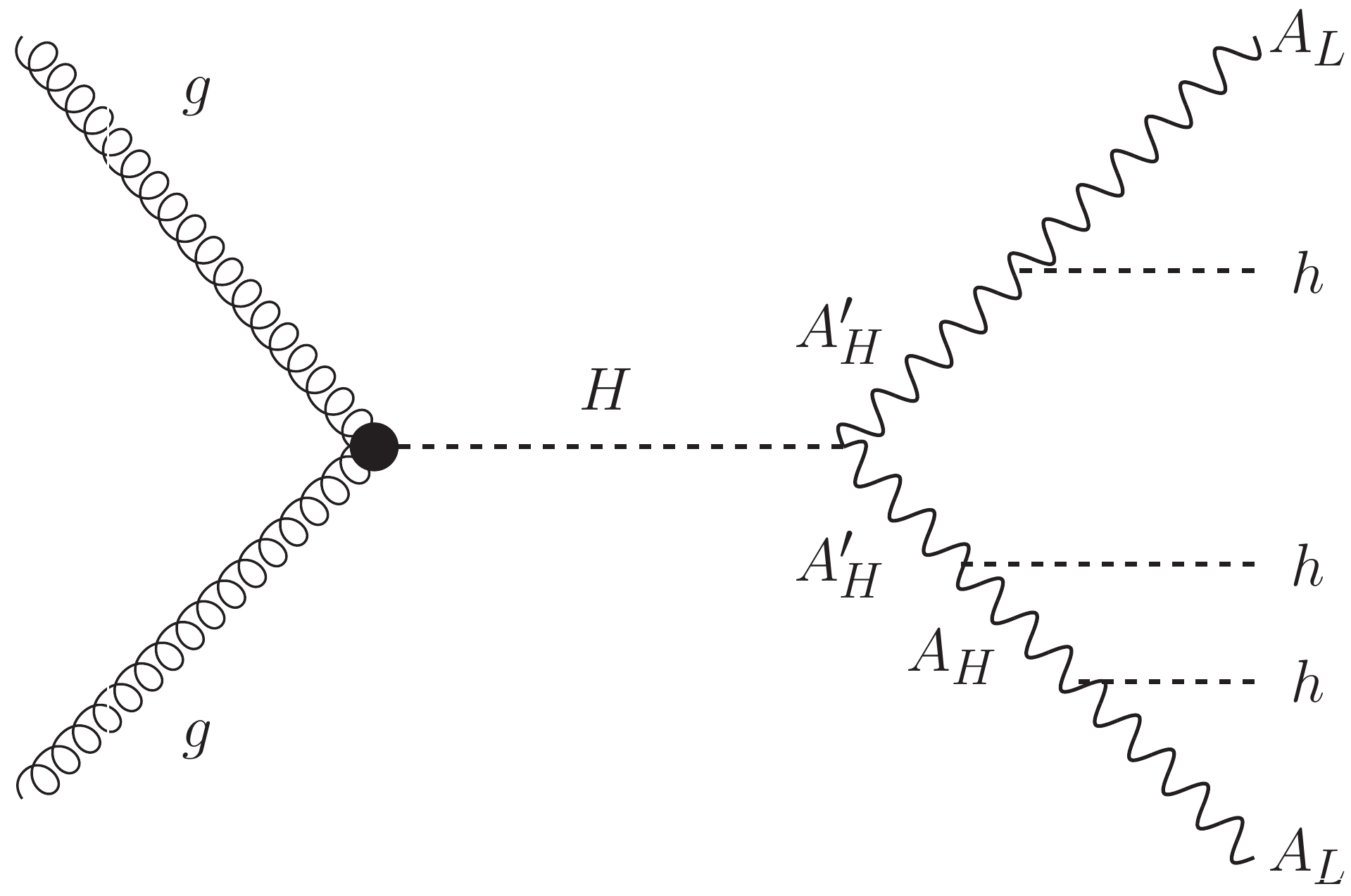} 
\caption{\underline{Left}: $h h$ + \cancel{$E$}$_T$ production.
\underline{Right}: $h h h$ + \cancel{$E$}$_T$ production.
\label{diags}}
\end{center}
\end{figure}

The di-Higgs + \cancel{$E$}$_T$ final state is generated through the diagram in 
 Fig.~\ref{diags}, left. We consider the regime where $H$ is produced on--shell and decays into a pair of $A_H$ with a significant branching fraction. These 
 subsequently decay into $A_L$'s and $h$'s as long as kinematically allowed, while 
 the $A_L$ pair escapes undetected thereby producing the missing energy signal.

\begin{figure} 
\begin{center}
\includegraphics[scale=0.45]{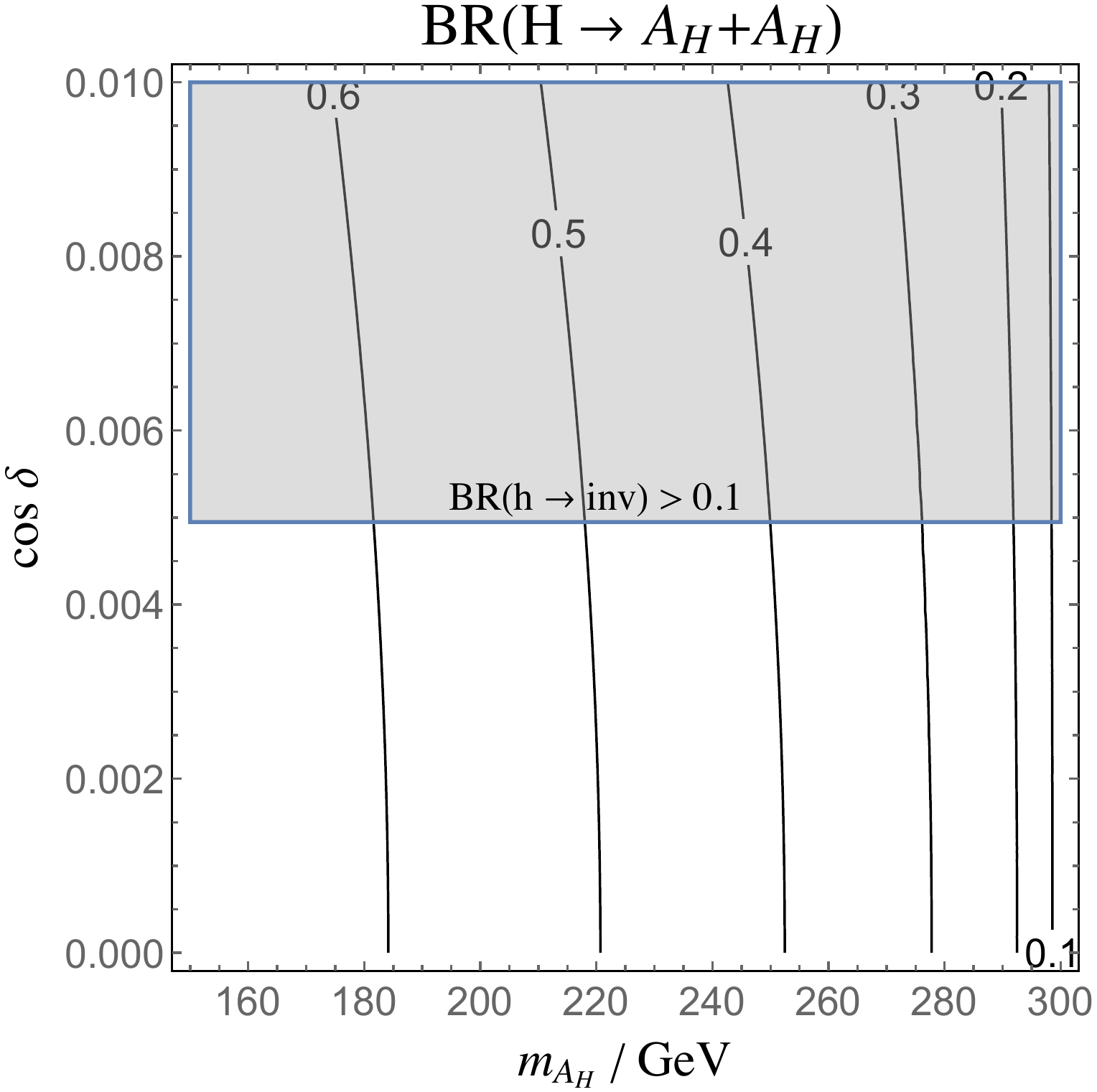} 
\caption{Branching ratio for the heavy Higgs decay to heavier dark bosons $H \to A_H A_H$.
Here $\sin \theta = 0.3$, $\tilde g = 2$, $m_H= 600$ GeV, $m_{A_L} = 10$ GeV, 
 $\BR(H \to h h)=0.2$.
The shaded region is excluded by $\BR(h \to \rm{inv})>0.1$.
\label{BRsimp}
}
\end{center}
\end{figure}

In the simplest model, $\BR(A_H \to A_L + h) \simeq 100\%$ for $m_{A_H} > m_{A_L}+m_h $. On the other hand, 
 $\BR(H \to A_H A_H)$ varies. 
The partial width for the $H$ decay into SM fermions and SM vector bosons is given by
\be
\Gamma(H \to f \bar f, Z Z, W^+ W^-) = \sin^2 \theta \times \Gamma_{\textrm{SM}}\vert_{m_h \to m_H} \,.
\ee
The partial widths for the $H$ decay to hidden vectors are given by
\besub
\bal
\Gamma(H \to A_L A_L) &= \frac{\tilde g^2 \cos^2 \theta \, \cos^2 \delta \, m_H^3}{128 \pi \, m_{A_L}^2} f (x_L) \;,
\\
\Gamma(H \to A_H A_H) &= \frac{\tilde g^2 \cos^2 \theta \, \sin^2 \delta \, m_H^3}{128 \pi \, m_{A_H}^2} f (x_H) \;,
\eal
\eesub
where 
$
x_{L/H} \equiv 4 m^2_{A_{L/H}}/{m_H^2} 
$
and 
$
f(x)= \left(1-x+ 3 x^2/4 \right) \sqrt{1-x} \,.
$
Fig.~\ref{BRsimp} shows the variation of $\textrm{BR}(H \to A_H A_H)$ with $m_{A_H}$ and $\cos\delta$.
The other parameters are set to $\sin \theta = 0.3$, $\tilde g = 2$, $m_H= 600$ GeV, $m_{A_L} = 10$ GeV and
 $\BR(H \to h h)$ is kept fixed at 0.2.
This choice is motivated by having a sizeable production cross section for $H$: both $\sin \theta$ and $\tilde g$
should be substantial, while $H$ should not be too heavy. We see that $\cos \delta$ is required to be small in order to get 
a significant $\textrm{BR}(H \to A_H A_H)$. A large portion of parameter space is excluded by the invisible $h$--decay constraint
$\BR(h \to \rm{inv})<0.1$, which also forces $\cos\delta \ll 1$. {More precisely, the relevant bound is that on the 
 signal strength $\mu$ 
of the 125 GeV Higgs which constrains a combination of $\sin\theta$ and the invisible decay BR (see e.g. the discussion in \cite{Huitu:2018gbc}). Given the fluctuations in the 
experimental values of $\mu$, we take the resulting constraint on $h \to \rm{inv}$ to be $\BR(h \to \rm{inv})<0.1$. The precise value of this bound
does not affect our results.}
In the SU(3) model, the smallness of $\cos\delta$ can be attributed to the Higgs mixing predominantly with one of the triplets, namely $\phi_1$,
such that $H$ has a large $\phi_1$ component
 (see Table 3 of \cite{Gross:2015cwa}.)
Finally, we note that the chosen values of $\sin\theta$ and $m_H$ are consistent with both the LHC and electroweak precision measurements \cite{Falkowski:2015iwa,Huitu:2018gbc}, although a   stronger bound on $\sin\theta$   from the $W$ mass alone      has been quoted in Ref.\,\cite{Robens:2015gla}.

 In this work, we are being agnostic as to the DM production mechanism and 
 do not impose a constraint on the thermal DM annihilation cross--section.\footnote{The thermal DM analysis for the SU(3) model has been performed in \cite{Gross:2015cwa}. } On the other hand, the direct DM detection cross--section is automatically suppressed: the $h$--mediated contribution is small due to $\cos^2\delta \ll 1$, while the $H$--mediated contribution is suppressed by $\sin^2 \theta $ and $m_H^4$. 
 Thus, in what follows we may focus on the collider aspects of the model as long as the parameter values are in the ballpark of those considered in this section.
 
The simplified model can be added a layer of complexity by including an extra heavy gauge boson $A_H^\prime$. Indeed, 
dark sectors with the symmetry group larger than SU(2) contain multiple heavy bosons with different masses. 
The relevant couplings take the form 
\begin{equation}
\mathcal{L} \supset \lambda_{abi} \; (A_a)_\mu (A_b)^\mu \, \Phi_i \;,
\end{equation}
where $A_a=A_H^\prime, A_H, A_L$ and $\Phi_i=h,H$.
Depending on the couplings $ \lambda_{abi}$ and masses, $H$ may predominantly decay into a pair of $A_H^\prime$, which then decay into $A_H$ and $A_L$ with $h$--emission. This can lead, for example, to a 3$h$ and dark matter final state as shown in Fig.~\ref{diags}, right. Such exotic processes may provide an additional handle on the dark sector properties.

 \section{LHC search for di--Higgs production with missing energy }

\subsection{ $2b+2\gamma + \slashed{E}_T$ in the final state}
\label{sec:coll}
Certain aspects of collider phenomenology of related models have been studied before 
 \cite{Djouadi:2011aa,Djouadi:2012zc,Chen:2015dea,Kim:2015hda,Kang:2015nga,Blanke:2019hpe,Adhikary:2017jtu,Alves:2019emf} through 
different production channels. In this work, we focus on the heavy CP-even scalar ($H$) production through gluon fusion, which subsequently 
decays via the hidden sector vector fields into the 125 GeV Higgs bosons ($h$) along with two dark matter particles. This can result in various 
final states depending on the decay mode of the 125 GeV Higgs boson. The dominant decay of the Higgs to
$b\bar b$ has been studied extensively in \cite{Blanke:2019hpe}. However, such a multi-$b$-jet final state is plagued with hadronic 
backgrounds and one needs a very clear understanding of the $W$+jets and QCD backgrounds at high luminosity in order to isolate the 
signal events. 
Using the jet substructure technique to reconstruct the Higgs bosons is an efficient tool for a heavy $H$, while in the mass range of interest to us it is less reliable.

Instead, we explore a cleaner channel where one of the 125 GeV Higgs bosons decays into a pair of photons. The photon identification 
efficiency is quite good and even though the corresponding branching ratio is small, an enhanced cut efficiency makes 
this channel a significant one. Thus, our signal region consists of two $b$-jets, two photons and large transverse missing energy. 
The LHC collaborations have studied the two $b$-jets + 2 photons signal region quite extensively in the context of BSM Higgs searches \cite{Sirunyan:2018iwt}. However,
this does not include large missing energy which we use as an additional feature to suppress the SM background.

The dominant SM background to the signal arises from the $t\bar th$, $b\bar bh$, $Zh$, $\gamma\gamma$+ jets, $t\bar t\gamma\gamma$, 
$b\bar b\gamma\gamma$ and $ZZ\gamma\gamma$ production. 
{ In addition, there are contributions from $b\bar b\gamma$ + jets and $b\bar b$ + jets with the jet(s) being mistagged as photon(s),
which can be important due to pile--up at the LHC at high luminosity.
 Although the mistagging probability is small, of order $ 0.1\%$ \cite{ATL-PHYS-PUB-2016-026}, the resultant background cross-section 
may not be negligible given the large production cross-section for  these two processes.} 
We have simulated all of the above  processes along with our signal for some representative 
benchmark points at the 14 TeV LHC. The parton level events have been generated using MadGraph5 \cite{Alwall:2011uj,Alwall:2014hca}.
We use the NNPDF parton distribution function \cite{Ball:2012cx, Ball:2014uwa} for our computation. 
These events are then passed through PYTHIA8 \cite{Sjostrand:2006za,Sjostrand:2014zea} for decay, showering and hadronisation.
For processes with additional jets at the parton level, MLM matching \cite{Hoche:2006ph,Mangano:2006rw} was performed through the MadGraph5-PYTHIA8 interface.
The complete event information is further passed through Delphes3 \cite{deFavereau:2013fsa,Selvaggi:2014mya,Mertens:2015kba} 
for detector simulation. The jets are constructed via FastJet \cite{Cacciari:2011ma} following the anti-kt \cite{Cacciari:2008gp} algorithm 
{ with radius parameter $R=0.4$}. 

\subsection{Problems with the cut-based analysis}
\label{sec:res_cut}

\begin{figure}[t]
\begin{center}
\includegraphics[scale=0.56]{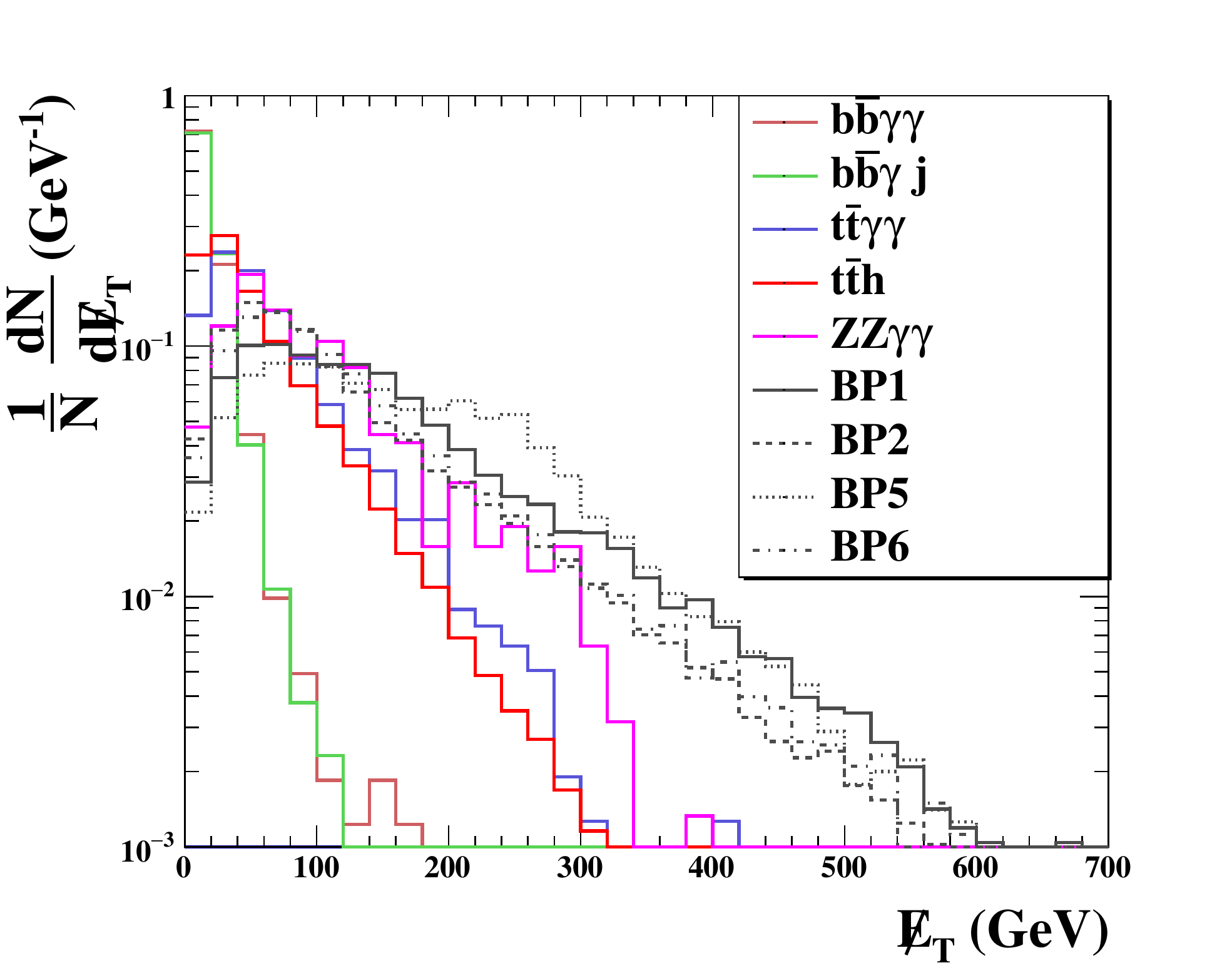}
\caption{ $\slashed{E}_T$ distribution for some of the  benchmark points and background processes for the final state 
$2b+2\gamma + \slashed{E}_T$.}
\label{fig:met_dist}
\end{center}
\end{figure}

{ In our  scenario,  the two $b$-jets and two photons in the final state arise from on-shell Higgs
 decays. In addition, the final state contains  two dark matter particles which 
 contribute  to the missing transverse energy. 
 In order to tag the $b$-jets, we have used a tagging efficiency which depends on the $b$ transverse momentum, 
 $0.80\times{\rm tanh}(0.003\times p_T)\frac{30}{(1+0.086\times p_T)}$ as provided by the ATLAS collaboration \cite{ATL-PHYS-PUB-2015-022}. 
 We have also considered the possibility of a light jet being mistagged as a $b$-jet. The mistagging probability is largest for the
 $c$-jets: $0.20\times{\rm tanh}(0.02\times p_T)\frac{1}{(1+0.0034\times p_T)}$ and much lower for the other light jets: 
 $0.002+7.3\times 10^{-6}\times p_T$ \cite{ATL-PHYS-PUB-2015-022}. In order to tag isolated charged leptons and photons, 
 we impose the requirements $p_T > 20$ GeV and  $|\eta| < 2.5$, where $\eta$ is the pseudorapidity. We further ensure that all the leptons are 
 well separated among themselves with $\Delta R_{\ell\ell} > 0.2$ and also from  other particles,  $\Delta R_{\ell x} > 0.4$, where $x$ can be 
 either photons or jets. Furthermore,  the transverse energy deposited by the hadrons within a cone of $\Delta R \le 0.2$ 
 around the  lepton is required  to be less than $0.2\times p_T^{\ell}$. Similar $\Delta R$ and hadronic energy deposition criteria are also imposed on the 
 photons for them to be considered  isolated. { A jet-jet separation $\Delta R_{jj} > 0.4$ is also imposed.}
We use the following kinematic cuts to achieve a good signal to background ratio:\footnote{Many of these cuts are borrowed from the CMS analysis \cite{Sirunyan:2018iwt}. }
\begin{itemize}[noitemsep]
\item {\bf C1:} The final state must contain two b-jets with $p_T^{b} > 25$ GeV and two photons.
Pseudorapidity of both the b-jets and the photons must lie within $|\eta| < 2.5$. 
There should be no isolated charged leptons in the final state.
\item {\bf C2:} The transverse missing energy $\slashed{E}_T$ must be larger than 120 GeV. { This cut is particularly 
useful to reduce background contributions from the channels with no direct source of missing energy, namely, $b\bar bh$, $Zh$, 
$\gamma\gamma$+ jets, $b\bar b\gamma\gamma$, $b\bar b\gamma$ + jets and $b\bar b$ + jets. For all these channels,  $\slashed{E}_T$ 
in the final state arises from  mismeasurement of the jet transverse momenta. Its distribution is likely to be much softer 
than that of   the  channels like $t\bar th$, $t\bar t\gamma\gamma$, and $ZZ\gamma\gamma$  as shown 
in Fig.~\ref{fig:met_dist}.} 
\item {\bf C3:} Invariant mass of the b-jet pair must lie within a 20 GeV window of the 125 GeV Higgs, $|m_{b\bar b}-125 \;{\rm GeV}| < 20$ GeV. 
{ The background  $b$-jets  are either hard produced  or arise from top quark or $Z$ boson decay. Hence 
the invariant mass of the $b$-jet pairs is  distributed over a wide range in most  cases, whereas for $Zh$ and $ZZ\gamma\gamma$ 
 it peaks  around the $Z$ boson mass as can be seen from Fig.~\ref{fig:minvbb_dist}. Thus, the cut on the $b \bar b $ invariant mass reduces the background contribution 
 drastically.}
\begin{figure}[t]
\begin{center}
\includegraphics[scale=0.56]{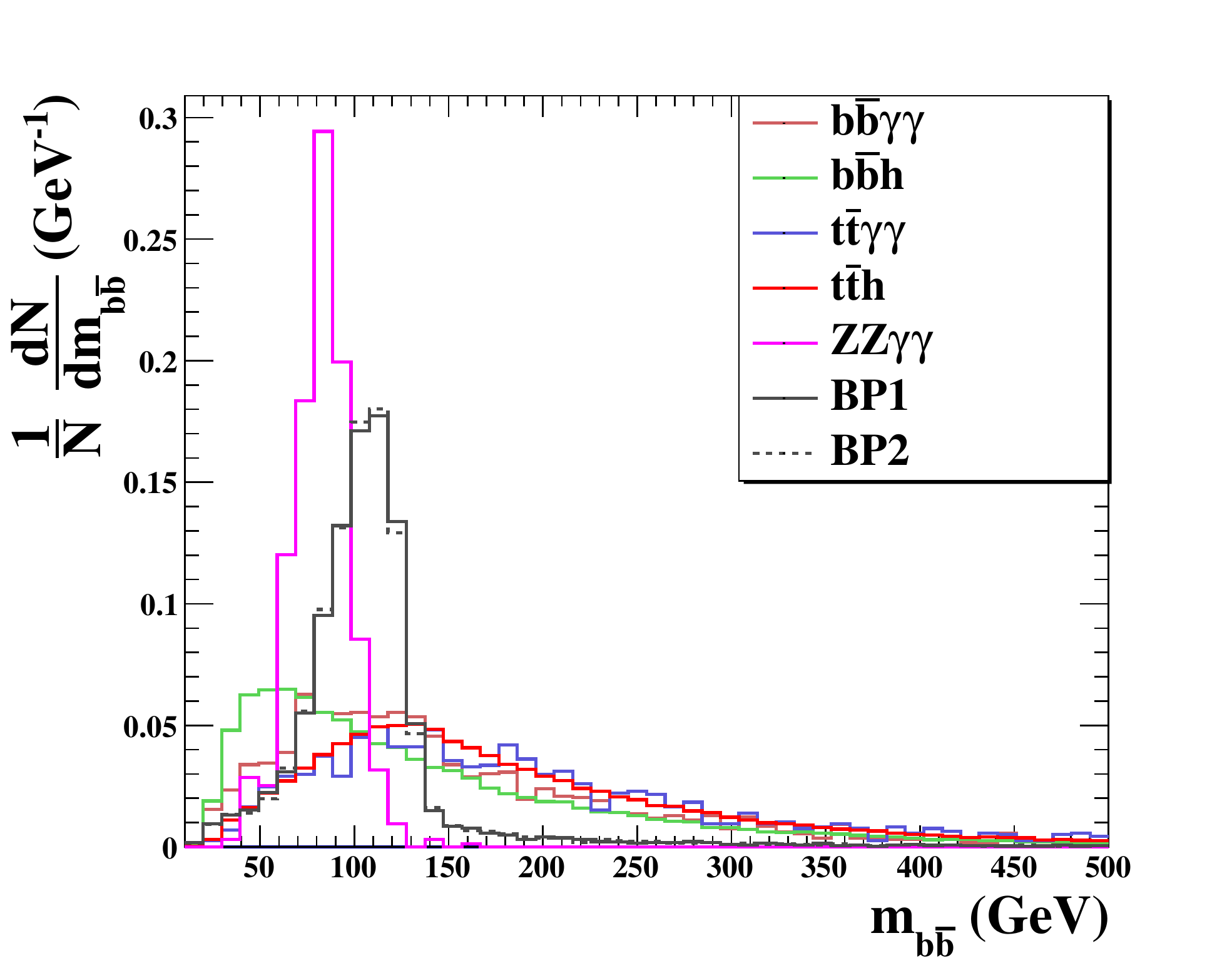}
\caption{ $m_{b\bar b}$ distribution  for some of the  benchmark points and background processes for the final state $2b+2\gamma + \slashed{E}_T$.}
\label{fig:minvbb_dist}
\end{center}
\end{figure}
\item {\bf C4:} Effective mass $m_{\rm eff}=\sum_i p_T^{b_i}+\sum_i p_T^{\gamma_i} + \slashed{E}_T $ must be larger than 800 GeV. 
For the signal events, this quantity is normally larger than that for the background channels like 
  $Zh$, $b\bar b h$. This cut 
also helps to reduce background contributions from  soft $b$-jets and photons in processes like $b\bar b\gamma\gamma$, 
$b\bar b\gamma$ + jets and $b\bar b$ + jets. 
\item {\bf C5:} Invariant mass of the photon pair must lie within a 5 GeV window of the 125 GeV Higgs, $|m_{\gamma\gamma}-125\;{\rm GeV}| < 5$ GeV.
The Higgs mass reconstruction from the photon pair is much more precise than that from the $b$-jets due to the better photon momentum resolution,
thus allowing for efficient discrimination against many of the background channels (except for $t\bar th$, $b\bar bh$ and $Zh$). The corresponding kinematic distribution for some of the  signal 
and background processes is shown in Fig.~\ref{fig:minvaa_dist}.
\begin{figure}[t]
\begin{center}
\includegraphics[scale=0.56]{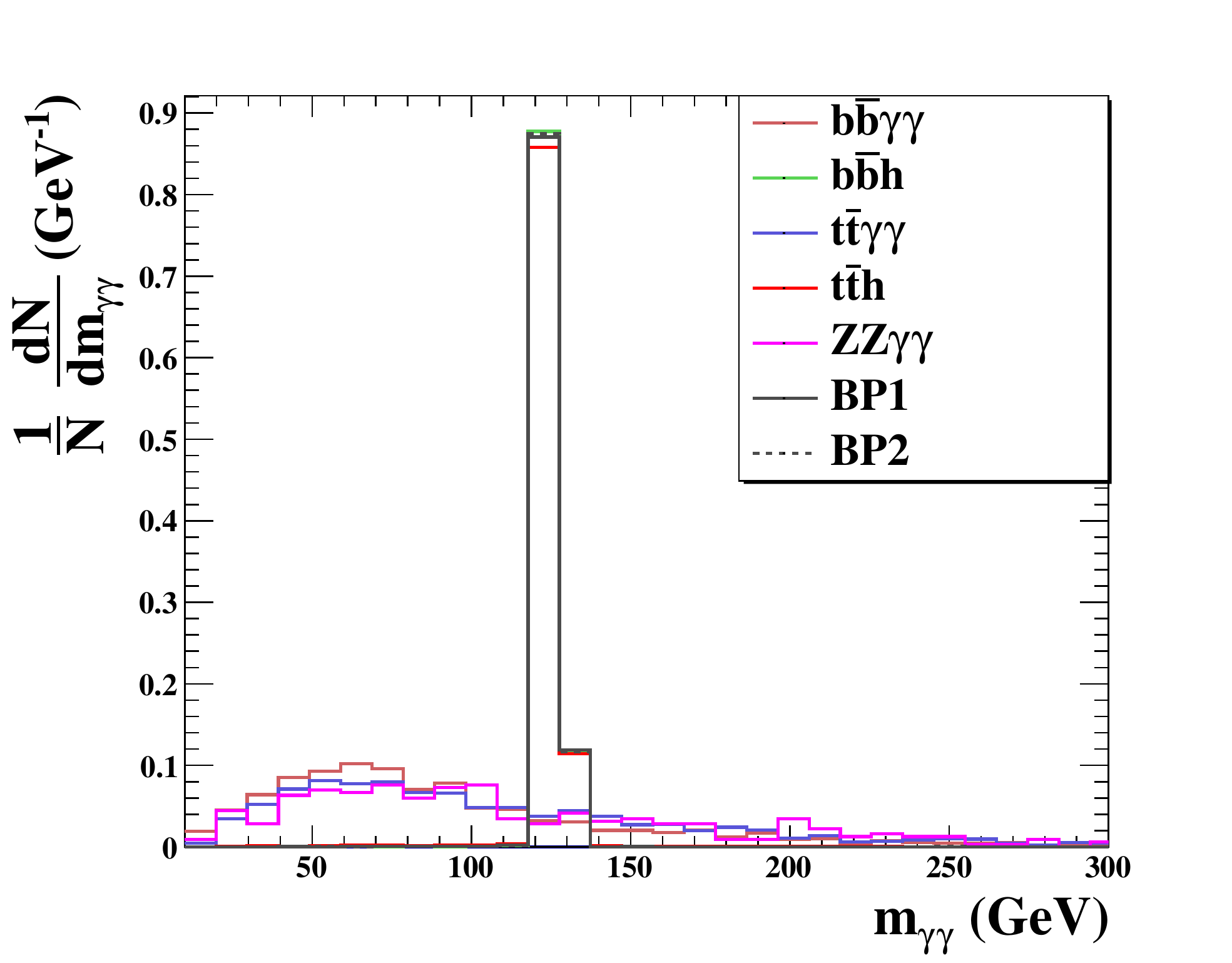}
\caption{Distribution of $m_{\gamma\gamma}$ for some sample benchmark points and background processes for the final state $2b+2\gamma + \slashed{E}_T$.}
\label{fig:minvaa_dist}
\end{center}
\end{figure}
\item {\bf C6:} Angular separation between the $b\bar b$ pair and the photon pair must be large, $\Delta\phi(b\bar b, \gamma\gamma) > 2.0$.
{ Since the two Higgs bosons are radiated from two different legs  in the production process, it is expected that the resultant $b$-jet 
pair and the $\gamma$ pair will be well separated in the final state. This is not  the case for some of the background processes, especially those 
where all the $b$-jets and photons are hard produced, like $b\bar b\gamma\gamma$, etc. The two $b$-jets can also arise from decays of two different 
mother particles, as in $t\bar t\gamma\gamma$, in which case the combined $b\bar b$ system may not be  well separated from the $\gamma\gamma$ 
system.}
\item {\bf C7:} $\frac{p_T^{\gamma_1}}{m_{\gamma\gamma}} > \frac{1}{3}$ and $\frac{p_T^{\gamma_2}}{m_{\gamma\gamma}} > \frac{1}{4}$. 
This eliminates some of the distortions  in the 
low end of the $m_{\gamma\gamma}$ distribution and reduces the continuum background.
\end{itemize} 
In Tables~\ref{tab:res_cut_bkd} and \ref{tab:res_cut_sig} we show our results obtained from the cut-based analysis. 
The signal strength depends strongly on $m_H$, $\sin\theta$ and BR($H\to A_H A_H$). For these we choose the benchmark values 
sin$\theta$=0.3, BR($H\to A_H A_H$) = 1.0 and 0.7 in order to maximize the event rate, while $m_H$ is taken to be 500 GeV  and above.
For lower $m_H$, the existing LHC data already constrain the model as we show in the subsequent section. Hence, $m_H \geq 500$ GeV
represents  the range of interest for the future LHC runs, as long as $\sin\theta$ and BR($H\to A_H A_H$) are relatively large. 
To paint a more complete picture, in Section 3.4.1 we 
 discuss the LHC reach in terms of $m_H$ leaving BR($H\to A_H A_H$) as a free parameter.

{Cuts {\bf C2}, {\bf C3} and 
{\bf C5} prove to be the most effective ones in reducing the background. 
\begin{table}[h]
\begin{center}
\begin{tabular}{||c||c|c||} 
\hline\hline
Channels & Production cross-section (fb) & Cross-section after cuts (fb) \\
\hline\hline
$t\bar th(h\to\gamma\gamma)$  &0.673 &$2.2\times 10^{-4}$  \\ 
$b\bar bh(h\to\gamma\gamma)$  &0.164 &$6.4\times 10^{-6}$  \\ 
$Zh(Z\to b\bar b, h\to\gamma\gamma)$ &0.141 & --  \\ 
$\gamma\gamma$+ jets  &61660.7 & --  \\ 
$t\bar t\gamma\gamma$  &2457.6 & $2.4\times 10^{-3}$   \\ 
$b\bar b\gamma\gamma$  &5144.41 &  --  \\
$ZZ(\to\nu\bar\nu b\bar b)$ + photons  &2.106 & -- \\ 
$b\bar b\gamma$ + jets &$9.76\times 10^6$  & -- \\
$b\bar b$ + jets   &$7.74\times 10^9$  & -- \\
\hline\hline
\end{tabular}
\caption{The SM background channels along with their cross-sections before and after the cuts. Numerically negligible results are not shown. {Here  we have used a flat rate of $0.1\%$ for the probability of a jet being mistagged as a photon.}}
\label{tab:res_cut_bkd}
\end{center}
\end{table} 
\begin{table}[h]
\begin{center}
\begin{tabular}{||c||c|c||} 
\hline\hline
Benchmark & Production & Cross-section \\
Points & cross-section (fb) & after cuts (fb) \\
\hline\hline
{\bf BP1:} & & \\ 
$m_H=500$ GeV, $m_{A_H}=240$ GeV, $m_{A_L}=10$ GeV &0.640 &$3.66\times 10^{-3}$   \\ 
{\bf BP2:} &(0.448) &($2.56\times 10^{-3}$) \\ 
$m_H=500$ GeV, $m_{A_H}=240$ GeV, $m_{A_L}=100$ GeV &  &$3.05\times 10^{-3}$   \\ 
& & ($2.14\times 10^{-3}$) \\
\hline\hline
{\bf BP3:} & & \\ 
$m_H=600$ GeV, $m_{A_H}=280$ GeV, $m_{A_L}=10$ GeV &0.286 &$9.25\times 10^{-4}$   \\ 
{\bf BP4:} &(0.200) &($6.48\times 10^{-4}$) \\ 
$m_H=600$ GeV, $m_{A_H}=280$ GeV, $m_{A_L}=150$ GeV&  &$8.95\times 10^{-4}$   \\ 
& & ($6.27\times 10^{-4}$)\\
\hline\hline
{\bf BP5:} & & \\ 
$m_H=700$ GeV, $m_{A_H}=340$ GeV, $m_{A_L}=10$ GeV &0.133 &$4.16\times 10^{-4}$   \\ 
{\bf BP6:} &(0.093) &($2.91\times 10^{-4}$) \\ 
$m_H=700$ GeV, $m_{A_H}=340$ GeV, $m_{A_L}=200$ GeV &  &$3.33\times 10^{-4}$   \\ 
& & ($2.33\times 10^{-4}$)\\
\hline\hline
\end{tabular}
\caption{The signal cross-section $\sigma(pp\to H\to A_H A_H\to h A_L h A_L\to b\bar b\gamma\gamma A_L A_L)$ before and after the cuts. For all the benchmark points, sin$\theta$=0.3, 
BR($H\to A_H A_H$)=1.0 or 0.7 (in parentheses), and BR($A_H\to A_L h$)=1.0. }
\label{tab:res_cut_sig}
\end{center}
\end{table} 
Clearly, the combined impact of the chosen cuts is adequate for our purposes}, yet they also decrease the signal cross-section to the 
extent that any possible signal could only be observed at high-luminosity LHC. 
The 14 TeV LHC is projected to accumulate 
an integrated luminosity of 3000 ${\rm fb}^{-1}$, which appears insufficient to obtain {even $3\sigma$ statistical} significance in most of the parameter space.\footnote{We 
use the following definition of statistical significance: ${\mathcal S}= S/ \sqrt{ S+B+\sigma_B^2} $,
where $S$ and $B$ are the number of signal and background events, respectively, and 
 the background uncertainty is taken to be negligibly small, $\sigma_B \ll B$.} 
  We have also tried to soften the cuts to increase the signal rate, but that eventually results in a worse 
signal to background ratio. {The best case scenario here is BP1, for which  the significance factor 
at 3000 ${\rm fb}^{-1}$ is $\sim 2.5\sigma$ if we assume $\sigma_B=0$.} Therefore, our conclusion is that imposing rectangular cuts does not lead to good discovery prospects.
\subsection{Multivariate analysis}
\label{sec:bdtans}
Evidently, the cut-based analysis is not sensitive enough to probe the present scenario at the 14 TeV LHC even at high luminosity. Thus, we next 
explore the possibility of improving the analysis with machine learning techniques, namely the Gradient Boosted Decision Trees (BDT) \cite{Chen:2016btl}.
This method of data analysis is being used quite extensively in LHC searches to good effect \cite{Baldi:2014kfa,Oyulmaz:2019jqr,Bhattacherjee:2019fpt,
Blanke:2019hpe,Bakhet:2015uca,Field:1996rw}. We have chosen the XGBoost \cite{Chen:2016btl} toolkit for the gradient boosting analysis.
Below we list the kinematic variables used in the decision trees (cf. \cite{Sirunyan:2018iwt}).
\begin{enumerate}[noitemsep]
 \item Number of $b$-jets, $N_b$.
 \item Number of photons, $N_{\gamma}$.
 \item Transverse momentum of the hardest b-jet, $p_T^{b_1}$.
 \item Transverse momentum of the second hardest b-jet, $p_T^{b_2}$.
 \item Transverse momentum of the hardest photon, $p_T^{\gamma_1}$.
 \item Transverse momentum of the second hardest photon, $p_T^{\gamma_2}$.
 \item Missing transverse energy, $\slashed{E}_T$.
 \item Invariant mass of the hardest b-pair, $M_{\rm inv}^{b\bar b}$. 
 \item Invariant mass of the hardest photon pair, $M_{\rm inv}^{\gamma\gamma}$. 
 \item Effective mass, $M_{eff}$, defined as the scalar sum of the transverse momenta of the jets, photons and $\slashed{E}_T$.
 \item Angular separation between the $b\bar b$ pair and the photon pair, $\Delta\phi(b\bar b, \gamma\gamma)$.
 \item Ratio of the hardest photon $p_T$ and di-photon invariant mass, $R_1 = \frac{p_T^{\gamma_1}}{M_{\rm inv}^{\gamma\gamma}}$.
 \item Ratio of the second hardest photon $p_T$ and di-photon invariant mass, $R_2 = \frac{p_T^{\gamma_2}}{M_{\rm inv}^{\gamma\gamma}}$.
\end{enumerate}
We have chosen 1000 trees, maximum depth 4 and learning rate 0.01 for our analysis. We combine data on the above kinematic 
variables for our signal events with all the background events in one data file. 
All the events are required to have at least two b-jets ($p_T^b > 25$ GeV), at least two photons ($p_T^{\gamma} > 20$ GeV) 
and no charged leptons with $p_T > 20$ GeV. We have combined the background events after properly weighting them according to their cross-sections subject 
to these cuts.
 As a result, more importance is given to the dominant backgrounds while training the data. We also make sure that there are 
enough signal events to match the total weight of the background events. We take $80\%$ of our data for training and 20$\%$ for testing.
{ Each of our data samples typically contains   values of the kinematic variables corresponding to $\sim 10^6$ events. 
These include the  final states subject to our conditions  on b-jets, photons and charged leptons as discussed above. In order to 
obtain this data set,  we have generated  $\sim 10^7$  Monte Carlo events  for the backgrounds and $\sim 10^6$ events for the signal benchmark points. }
\subsubsection{Results }
\label{sec:res_bdt}
Fig.~\ref{fig:bdt_response} shows results of our multivariate analysis for benchmark point BP1. 
The BDT classifier response shows clear distinction between the signal and the background. We find that among the kinematic 
variables listed above, $M_{\rm inv}^{b\bar b}$ and $\slashed{E}_T$ are the two most important discriminators followed by $M_{\rm inv}^{\gamma\gamma}$.
The right panel of Fig.~\ref{fig:bdt_response} shows 
 the Receiver Operating Characteristic (ROC) curve for BP1. 
The x-axis (efficiency) indicates the fraction of identified signal events after imposing the BDT classifier while the y-axis (Purity) 
indicates the ratio of identified signal events to the total number of identified events (signal plus background) after imposing the classifier. 
The area under the curve (AUC) is a good indicator of the BDT performance. For this benchmark point, AUC=0.90.
\begin{figure}[h]
\begin{center}
\includegraphics[scale=0.46]{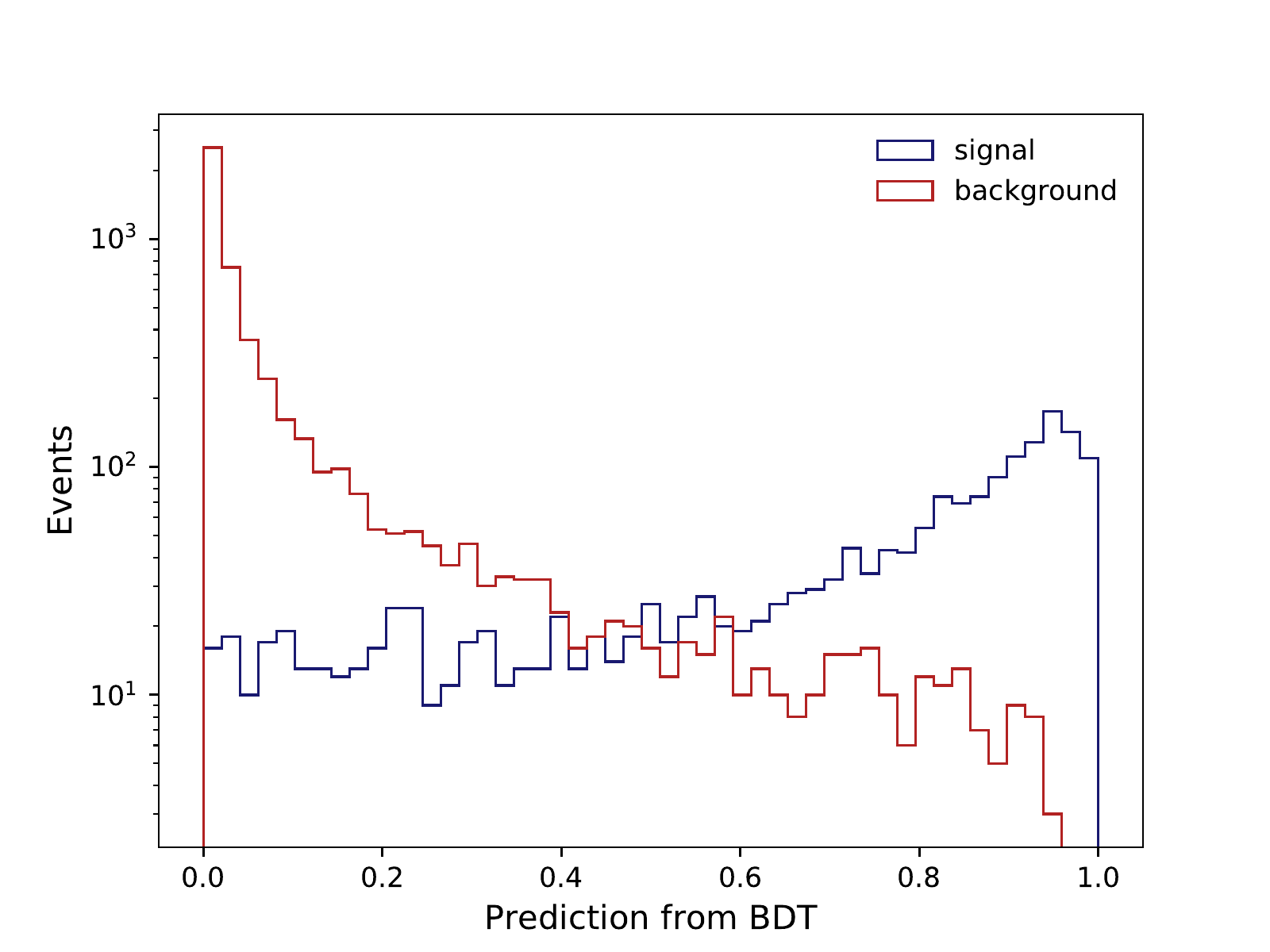}
\includegraphics[scale=0.46]{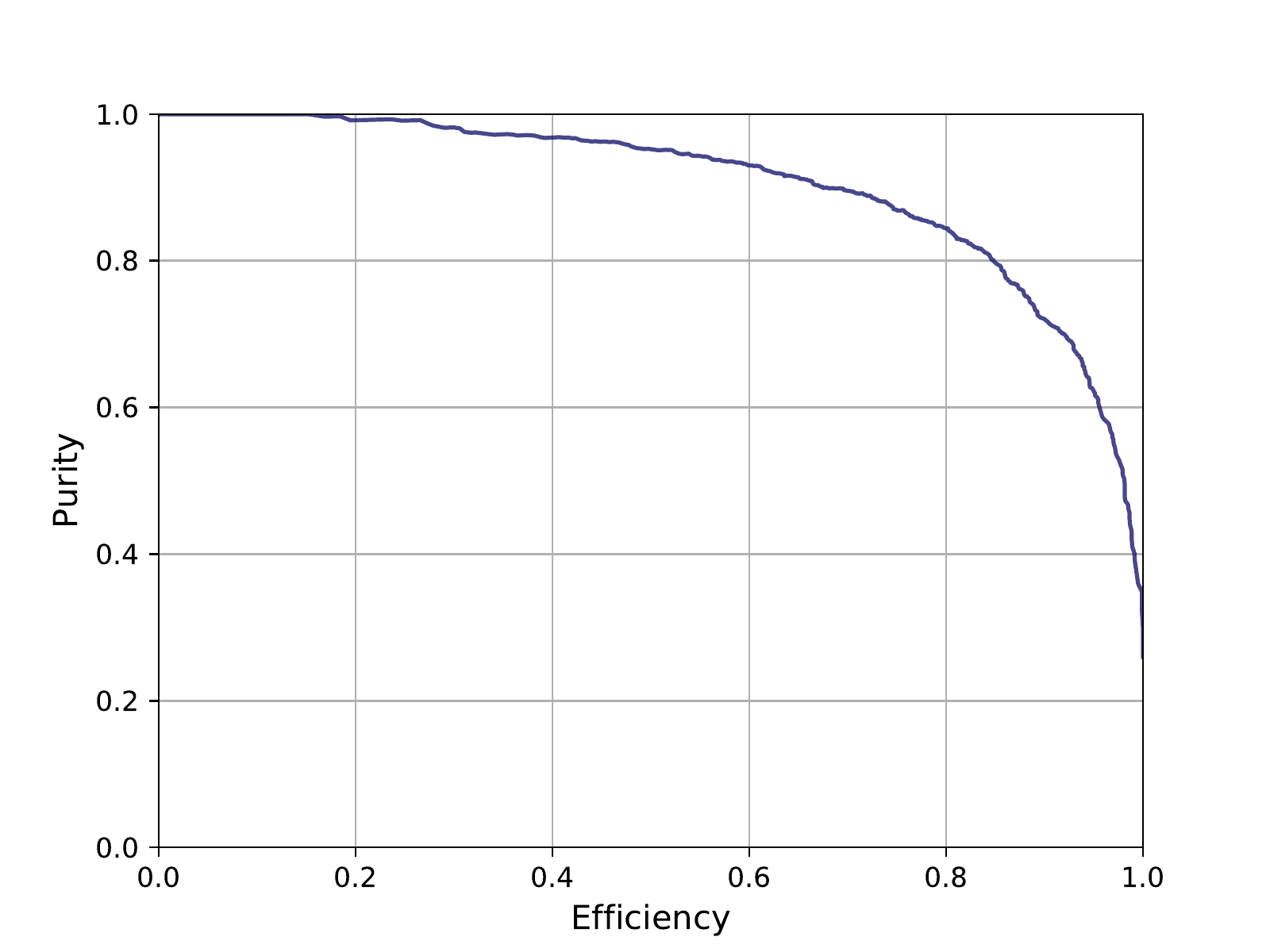}
\caption{Example of the BDT classifier response for BP1 ($m_H=500$ GeV, $m_{A_H}=240$ GeV, $m_{A_L}=10$ GeV) in differentiating the signal 
and background events. The ROC curve in the right panel indicates efficiency of the algorithm.}
\label{fig:bdt_response}
\end{center}
\end{figure}
 
Our multivariate analysis results are quite promising and prove to be a significant improvement over those for the cut-based analysis. 
The AUC indicator is close to 0.9 for all the benchmark points, which shows that the BDT classifier is efficient in 
distinguishing the signal from the background in all considered cases.
 In Table~\ref{tab:res_bdt}, we present the AUC for the benchmark points along with the required 
luminosity to achieve a signal significance of 3$\sigma$. 
\begin{table}[b]
\begin{center}
\begin{tabular}{||c||c|c||} 
\hline\hline
Benchmark & Area under the & Required \\
points & ROC curve & luminosity (${\rm fb}^{-1}$) \\
\hline\hline
{\bf BP1} &0.89 &300 (505) \\ 
{\bf BP2} &0.87 &707 (1264) \\ 
{\bf BP3} &0.91 &870 (1433) \\ 
{\bf BP4} &0.88 &2255 (4174) \\ 
{\bf BP5} &0.91 &3911 (7381) \\ 
{\bf BP6} &0.88 &5995 (11469) \\ 
\hline\hline
\end{tabular}
\caption{The area under the ROC curve for the benchmark points (see Table\;\ref{tab:res_cut_sig}) along with the required luminosity to achieve a 3$\sigma$ signal significance in the $2b+2\gamma + \slashed{E}_T$ channel at the 14 TeV LHC assuming BR($H\to A_H A_H$)=1 or 0.7 (in parentheses).
 }
\label{tab:res_bdt}
\end{center}
\end{table} 

This approach allows one to constrain the low $H$--mass window already with the existing LHC data. 
The integrated luminosity required to achieve the $3\sigma$
significance  for $m_H =  450$ GeV, $m_{A_H} = 150$ GeV, sin$\theta$ = 0.3 and  BR($H\to A_H A_H$) = 1.0, 0.7
  is 72 and 114 ${\rm fb}^{-1}$, respectively.  Thus, the current set of 139 ${\rm fb}^{-1}$  of data collected should be sufficient 
  to exclude $m_H \leq  450$ GeV for these parameter values, although the $2b+2\gamma + \slashed{E}_T$ channel
  has not yet been analyzed by ATLAS and CMS. 
  This justifies our choice of the $m_H$ benchmark values of 500 GeV and above for the future LHC runs.

{From our BDT analysis, we conclude 
that 9 out of the 13 kinematic variables used are essential to obtain the sensitivity indicated in Table~\ref{tab:res_bdt}. The four less essential variables are $N_b$, $N_{\gamma}$, $R_1$ and $R_2$. Among these, $N_b$ and $N_{\gamma}$ are rendered inconsequential by the 
criteria we have set  to select the final states, which ensure that all the signal and background events already have at least one pair of b-jets and photons. 
Some of the important kinematic variables, specifically $p_T^{\gamma_1}$, $p_T^{\gamma_2}$ and $M_{\rm inv}^{\gamma\gamma}$, ensure that $R_1$ and $R_2$ are not essential either. With 
the remaining 9 variables, we have performed a 10-fold cross validation check in order to assess the robustness of our analysis. We have randomly 
split our data set in train and test sets of similar size and computed the signal significance for 10 such different combinations. In Table~\ref{tab:10fold_bbaamet}, we present the   resulting signal significance factors obtained at an integrated luminosity of $1~{\rm fb}^{-1}$ along with the corresponding mean and the variance for the  benchmark point BP1. The variance is clearly quite small, indicating the robustness of our results.}
\begin{table}[t]
\begin{center}
\begin{tabular}{||c|c|c||} 
\hline\hline
Significance factors & Mean & Variance \\
\hline\hline
0.171, 0.174, 0.172, 0.168, 0.175, & & \\
0.171, 0.176, 0.178, 0.169, 0.172 & 0.1726 & $3.0\times 10^{-3}$ \\
\hline\hline
\end{tabular}
\caption{Significance factors, their mean and their variance  obtained in our 10-fold validation procedure for $2b+2\gamma + \slashed{E}_T$ at 
BP1.  The  integrated  luminosity is taken to be  $1~{\rm fb}^{-1}$.}
\label{tab:10fold_bbaamet}
\end{center}
\end{table} 
\subsection{Multivariate analysis for $2 b$--jets + $2 \ell + \slashed{E}_T$ in the final state}
\label{sec:multv_2b2lmet}
Another promising channel for the di--Higgs and dark matter search is provided by the $\bar b b $ and $WW^*$ decay modes of the Higgs bosons. 
Tagging on leptonic decays of the $W$'s, one obtains a clean final state with 2$b$--{jets} $ + 2\ell + \slashed{E}_T$. Here we take $\ell = e , \mu$.
The challenge is to suppress the $t\bar t + {\rm jets}$ background which has a much larger cross-section. 
Further background contributions come from the $t\bar t V$, $t\bar t h$, $VVV$ and $VV + {\rm jets}$ 
($V\equiv W^{\pm}, Z$) channels. We perform a multivariate analysis of our signal with the following kinematic variables (cf. \cite{Aad:2019yxi}): 
\begin{enumerate}[noitemsep]
\item Number of $b$-jets, $N_b$.
\item Number of charged leptons ($e$ or $\mu$), $N_{\ell}$.
\item Number of non-b-tagged jets, $N_j$.
\item Transverse momentum of the hardest $b$-jet, $p_T^{b_1}$.
\item Transverse momentum of the second hardest $b$-jet, $p_T^{b_2}$.
\item Transverse momentum of the hardest charged lepton, $p_T^{\ell_1}$.
\item Transverse momentum of the second hardest charged lepton, $p_T^{\ell_2}$.
\item Missing transverse energy, $\slashed{E}_T$. 
\item Effective mass defined as a scalar sum of the transverse momenta of the b-jets, leptons and $\slashed{E}_T$.
\item Invariant mass of the $b$-jet pair, $m_{\rm inv}^{b\bar b}$.
\item Invariant mass of the lepton pair, $m_{\rm inv}^{\ell\bar\ell}$.
\item Transverse momentum of the dilepton system, $p_T^{\ell\bar\ell}$.
\item Transverse momentum of the $b$-jet pair, $p_T^{b\bar b}$. 
\item Separation between the two charged leptons, $\Delta R^{\ell\bar\ell}$.
\item Separation between the two $b$-jets, $\Delta R^{b\bar b}$.
\item Azimuthal separation between the $b$-jet pair and the dilepton system, $\Delta\phi(b\bar b, \ell\bar\ell)$.
\end{enumerate}

\subsubsection{Results}

We collect information on these kinematic variables subject to the following preliminary cuts: the final state is required to have at least two 
$b$-jets with $p_T > 25$ GeV, at least two charged leptons with $p_T > 20$ GeV and $\slashed{E}_T > 50$ GeV. 
The final results are summarised in Table~\ref{tab:res_bdt_2b2lmet}. 
\begin{table}[h]
\begin{center}
\begin{tabular}{||c||c|c||}
\hline\hline
Benchmark & Area under the & Required \\ 
Points & ROC curve & Luminosity (${\rm fb}^{-1}$) \\
\hline\hline
{\bf BP1} &0.88 &463 (900) \\
{\bf BP2} &0.87 &819 (1614) \\
{\bf BP3} &0.85 &2900 (5804) \\
{\bf BP4} &0.83 &11468 (23192) \\
\hline\hline
\end{tabular}
\caption{ The area under the ROC curve for the benchmark points (see Table\;\ref{tab:res_cut_sig}) along with the required integrated luminosity to achieve a 3$\sigma$ signal significance
in the $2b+2\ell + \slashed{E}_T$ 
at the 14 TeV LHC assuming BR($H\to A_H A_H$)=1 or 0.7 (in parentheses). BP5 and BP6 require very large integrated luminosity and thus not shown. }
\label{tab:res_bdt_2b2lmet}
\end{center}
\end{table}
In Fig.~\ref{fig:disc_bp}, we present the required integrated luminosity
for a 3$\sigma$ signal significance in both channels.
\begin{figure}[t]
\begin{center}
\includegraphics[scale=0.335]{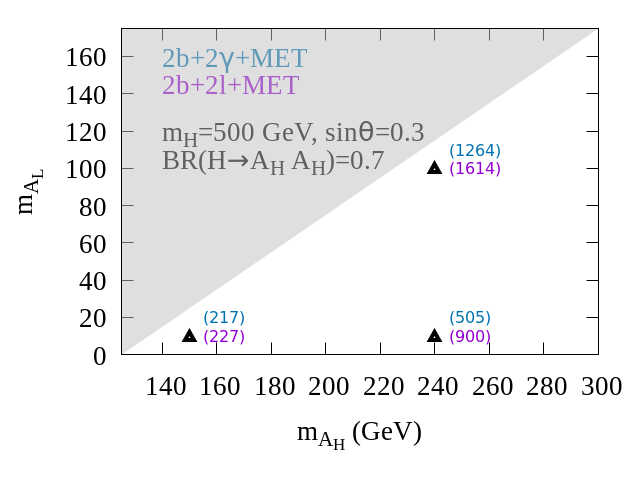}
\includegraphics[scale=0.335]{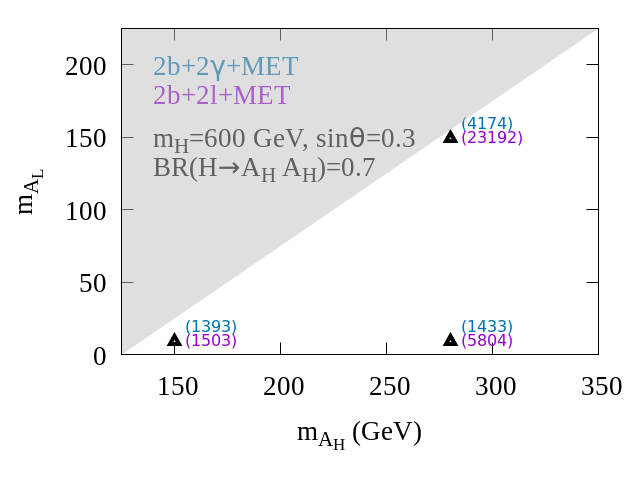}
\caption{Integrated luminosity required for a $3\sigma$ signal significance in 
 the $2b+2\gamma + \slashed{E}_T$ and $2b+2\ell + \slashed{E}_T$
channels at 14 TeV LHC.}
\label{fig:disc_bp}
\end{center}
\end{figure}

{ Apart from $N_b$ and $N_{\ell}$, all the  kinematic variables are essential  with $m_{\rm inv}^{b\bar b}$, $\slashed{E}_T$, $m_{\rm inv}^{\ell\bar\ell}$ and $\Delta R^{b\bar b}$ being the most important ones. We have performed a 10-fold cross validation check for this analysis as well following the 
prescription  of Section~\ref{sec:res_bdt}. In Table~\ref{tab:10fold_bbllmet}, we present the resulting 10  signal significance factors obtained at an integrated luminosity of $1~{\rm fb}^{-1}$ along with the corresponding mean and the variance for the benchmark point BP1. The variance is sufficiently small  indicating  robustness of our results.}
\begin{table}[b]
\begin{center}
\begin{tabular}{||c|c|c||} 
\hline\hline
Significance factors & Mean & Variance \\
\hline\hline
0.143, 0.144, 0.140, 0.138, 0.143, & & \\
0.137, 0.136, 0.136, 0.139, 0.138 & 0.1394 & $2.8\times 10^{-3}$ \\
\hline\hline
\end{tabular}
\caption{  Significance factors, their mean and their variance  obtained in our 10-fold validation procedure for $2b+2\ell + \slashed{E}_T$  at 
BP1.  The  integrated  luminosity is taken to be  $1~{\rm fb}^{-1}$.  }
\label{tab:10fold_bbllmet}
\end{center}
\end{table} 

The results can also presented in terms of the exclusion limits on the model. In particular, the branching ratio for the decay $H\rightarrow A_H A_H$ can be severely constrained with 3ab$^{-1}$ of data. Fig.~\ref{fig:disc_2h} displays the corresponding bound for a representative set of parameters, namely 
$m_{A_H}=150$ GeV, $m_{A_L}=10$ GeV and ${\rm sin}\theta=0.2,0.3$. The constraint can be as strong as BR($H\to A_H A_H$)$<7$\% for $m_H < 400$ GeV. 
Substantial values of BR($H\to A_H A_H$) can be probed up to $m_H \sim 600-700$ GeV. We also see that the $2b+2\gamma + \slashed{E}_T$ channel performs slightly better than 
 $2b+2\ell + \slashed{E}_T$ does.

\begin{figure}[t]
\begin{center}
\includegraphics[scale=0.40]{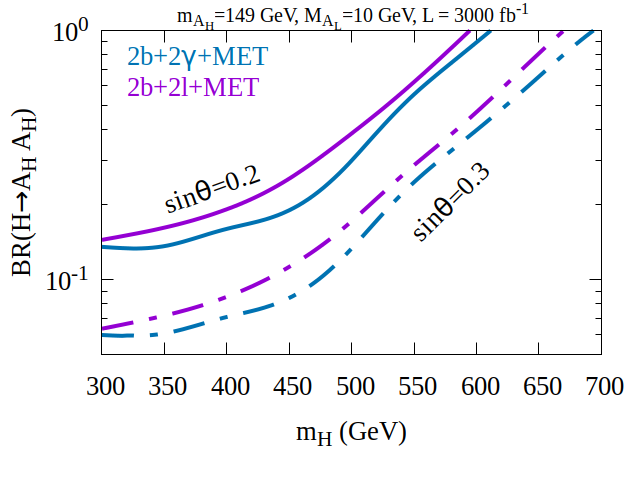}
\caption{ $3\sigma$ sensitivity of the $2b+2\gamma + \slashed{E}_T$ and $2b+2\ell + \slashed{E}_T$
channels to ${\rm BR}(H\to A_H A_H)$ at 14 TeV LHC with integrated 
luminosity of $3000~{\rm fb}^{-1}$. The dashed (solid) line corresponds to ${\rm sin}\theta=0.3$ (${\rm sin}\theta=0.2$).}
\label{fig:disc_2h}
\end{center}
\end{figure}
 
{ So far, we have presented our results assuming  $\sigma_B\simeq 0$. It is very difficult to forecast the future background uncertainty,
so we have to  resort to simple estimates. 
Let us 
 assume the background uncertainty to be $20\%$ of the total background, i.e., 
$\sigma_B=0.2\times B$. Consequently,   $\sigma_B^2$ can be  very large   for signal 
regions with a large background cross-section. The    $2b+2\ell + \slashed{E}_T$   signal is  thus affected quite  severely 
since $t\bar t$ + jets is a direct background to this final state and it comes with a large irreducible cross-section. We find that 
at $3000~{\rm fb}^{-1}$
luminosity with the parameter choices shown in Fig.~\ref{fig:disc_2h}, the signal significance  for the $2b+2\gamma + \slashed{E}_T$ final 
state reduces by a factor of $\sim 2$ when $\sigma_B=0.2\times B$ compared to the case with $\sigma_B=0$, whereas for $2b+2\ell + \slashed{E}_T$ 
it reduces by $\sim 10$. Since at such high luminosity, ${\mathcal S}= S/ \sqrt{ S+B+\sigma_B^2} \simeq S/ \sqrt{ B+\sigma_B^2} $, the required 
${\rm BR}(H\to A_H A_H)$ in the $2b+2\gamma + \slashed{E}_T$ and $2b+2\ell + \slashed{E}_T$ final states have to be scaled up by factors of 2 and 10, 
respectively. The scale factor remains the same for the different choices of ${\rm sin}\theta$ since  the kinematics of the final state are not affected.} 
\section{LHC search for a tri--Higgs and dark matter final state}

\subsection{Multivariate analysis for $\ge 3 b$--{jets} $+ 2\ell + \slashed{E}_T$ in the final state}
\label{sec:multv_3b2lmet}
In the presence of additional heavy particles in the dark sector, more exotic final states can be produced. In particular, cascade decays can lead to 3 or more Higgs bosons $h$ in the final state.
{ For example, the heavy Higgs $H$ can produce  a pair of heavy $A_H^\prime$, which in turn decay  via two different decay modes $A_H^\prime\to A_H h$ and 
$A_H^\prime\to A_L h$. Subsequently, $A_H$ decays into $A_L$ and $h$, thereby generating a multi--Higgs final state.}
The channel with four Higgs bosons 
suffers from severe kinematic suppression, while the tri--Higgs one could potentially be interesting.

Multi--Higgs production can be efficient if the decays happen on--shell. This implies that $A_H^\prime$ must be heavier than 250 GeV and thus $m_H >500$ GeV.
Given a tri--Higgs final state, there are a number of options to consider for Higgs decay. If all three decay into $\bar b b$ pairs, the signal extraction is marred by a large QCD multi--jet background
and possible misidentification of light flavor jets as $b$--jets. For the current study, we choose instead 2 $\bar b b$ pairs and leptonically decaying $WW^*$ as our final state. 
It has the advantage of being relatively clean and, in addition, we may recycle many of the background calculations done in the previous section.
We perform a multivariate analysis with the following kinematic variables: 
\begin{enumerate}[noitemsep]
\item Number of $b$-jets, $N_b$.
\item Number of charged leptons ($e$ or $\mu$), $N_{\ell}$.
\item Number of non-b-tagged jets, $N_j$.
\item Transverse momentum of the hardest $b$-jet, $p_T^{b_1}$, involved in a b-jet pair with $m_{\rm inv}^{b\bar b}$ closest to 125 GeV.
\item Transverse momentum of the second hardest $b$-jet, $p_T^{b_2}$, involved in a b-jet pair with $m_{\rm inv}^{b\bar b}$ closest to 125 GeV.
\item Transverse momentum of the hardest charged lepton, $p_T^{\ell_1}$.
\item Transverse momentum of the second hardest charged lepton, $p_T^{\ell_2}$.
\item Missing transverse energy, $\slashed{E}_T$. 
\item Effective mass, defined as a scalar sum of the transverse momenta of the b-jets, leptons and $\slashed{E}_T$.
\item Invariant mass of the $b$-jet pair, $m_{\rm inv}^{b_1 b_2}$.
\item Invariant mass of the lepton pair, $m_{\rm inv}^{\ell\bar\ell}$.
\item Transverse momentum of the dilepton system, $p_T^{\ell\bar\ell}$.
\item Transverse momentum of the $b$-jet pair, $p_T^{b_1b_2}$. 
\item Separation between the two charged leptons, $\Delta R^{\ell\bar\ell}$.
\item Separation between the two $b$-jets, $\Delta R^{b_1b_2}$.
\item Azimuthal separation between the $b$-jet pair and the dilepton system, $\Delta\phi(b_1 b_2, \ell\bar\ell)$.
\end{enumerate}

We collect information on these kinematic variables imposing the following preliminary cuts: the final state is required to have at least 3
$b$--jets with $p_T > 25$ GeV and at least two charged leptons with $p_T > 20$ GeV. 
We construct all possible $b$--jet pairs in the final state, compute the corresponding invariant masses, and then identify the pair that has 
$m_{\rm inv}^{b\bar b}$ closest to 125 GeV. We call the harder $b$--jet in this pair $b_1$ and the other one is identified as $b_2$.

Note that we are not using the third $b$-jet kinematics directly in our multivariate analysis. Including it adds a few more variables to the list: $p^T_{b_3}$, 
$\Delta\phi(b_3, \ell\bar\ell)$ and $\Delta\phi(b_1 b_2, b_3)$. However, we have verified that these do not improve our results.

The sensitivity of this signal region is weaker compared to the previous two. Although the BDT classifier is quite efficient in isolating the signal 
events, the few remaining background events have a large enough cross-section to suppress the signal significance. This is due to the small signal 
cross--section to start with, which is further reduced by detector simulations and applying the BDT classifier. 

To give an example, consider the parameter set $m_H=600$ GeV, $m_{A_H^\prime}=290$ GeV, $m_{A_H}=150$ GeV, $m_{A_L}=20$ GeV. 
The production cross section for the required final state is $0.107$ fb. Optimizing the signal to background ratio using our 
multivariate analysis, we find that the 1$\sigma$ signal significance is achieved with 3000 fb$^{-1}$ integrated luminosity, 
while that at 2$\sigma$ level requires 12000 fb$^{-1}$. Clearly, this is beyond the reach of the LHC.
 
Although our result for the $\ $ $ 3 b$--{jets} $+ 2\ell + \slashed{E}_T$ channel is negative, a fully $b$--jet $+ \slashed{E}_T$ final 
state may be more promising. It requires a dedicated analysis which we reserve for future work. 
\section{Summary and Conclusions}
{In this work, we have  considered dark sectors with spontaneously broken gauge symmetries, where} dark cascade decays   naturally lead to multi--Higgs final states with missing energy.
We have introduced a simplified model which captures main features of realistic hidden sectors   containing dark matter as well as further heavier states.

The focus of this work is  2 and 3 Higgs final states which subsequently decay into $\bar b b, \gamma \gamma$ and $WW$.
Using multivariate analysis with Boosted Decision Trees, we find that the
$2b+2\gamma + \slashed{E}_T$ and $2b+2\ell + \slashed{E}_T$ channels are promising in the context of 14 TeV LHC with
3 ab$^{-1}$ integrated luminosity. In particular, light dark matter $A_L$ with mass $\lesssim 100$ GeV can be probed efficiently
for the dark Higgs ($H$) mass below 700 GeV and its mixing angle with the SM Higgs $\sin\theta \sim 0.2-0.3$. In this region,
3$\sigma$ and higher signal significance can be achieved. The result can also be translated into a bound on the dark Higgs decay into
the heavier partners of dark matter $A_H$, with sensitivity to ${\rm BR}(H\to A_H A_H)$ reaching 6\% in the best case scenario.

The 3 Higgs final state, on the other hand, appears far less promising, at least for the decay channels considered. Fully hadronic Higgs 
decays may change this situation, but require a dedicated background study and detection simulation.

\vspace{10pt}
\noindent
{\bf Acknowledgements} 

\noindent
M.F. is supported by the National Research Foundation of South Africa - The World Academy of Sciences (NRF-TWAS) Grant No. 110790, Reference No. SFH170609238739.
C.G. is supported by the European Research Council grant NEO-NAT.

{}

\end{document}